\documentstyle[epsfig]{aipproc}

\begin{document}
%
\title{Dynamical Bar Instability \break
in Relativistic Rotating Stars}
%
\author
{Motoyuki Saijo $^{*}$,
Masaru Shibata $^{* \dag}$,\\
Thomas W. Baumgarte $^{*}$ and~ 
Stuart L. Shapiro $^{* \ddag}$}
%
\address{
{$^{*}$}
Department of Physics, University of Illinois at Urbana-Champaign,\\
1110 West Green Street, Urbana, Illinois 61801-3080\\
{$^{\dag}$}
Department of Earth Science and Astronomy, 
Graduate School of Arts and Science, \\
University of Tokyo,
3-8-1 Komaba, Meguro, Tokyo 153-8902, Japan\\
{$^{\ddag}$}
Department of Astronomy and NCSA, University of Illinois at
Urbana-Champaign, \\
1002 West Green Street, Urbana, Illinois 61801}
%
\maketitle
%
\begin{abstract}
We study by computational means the dynamical stability against
bar-mode deformation of rapidly and differentially  rotating stars in
a post-Newtonian approximation of  general relativity.  We
vary the compaction of the star $M/R$  (where $M$ is the gravitational
mass and $R$ the equatorial  circumferential radius) between 0.01 and
0.05 to isolate the  influence of relativistic gravitation on the
instability.  For  compactions in this moderate range, the critical
value of $\beta = T/W$ for the onset of the dynamical instability
(where $T$  is the rotational kinetic energy and W the gravitational
binding  energy) slightly decreases from $\sim$ 0.26 to $\sim$ 0.25
with increasing compaction for our choice of the differential
rotational law.  Combined with our earlier findings based on
simulations in full general relativity for stars with higher
compaction, we  conclude that relativistic gravitation enhances the
dynamical  bar-mode instability, i.e. the onset of instability sets in
for  smaller values of $\beta$ in relativistic gravity than in Newtonian 
gravity.  We also find that once a triaxial structure forms after  the
bar-mode perturbation saturates in dynamically unstable  stars, the
triaxial shape is maintained, at least for several  rotational periods. 
To check the reliability of our numerical  integrations, we verify that
the general relativistic Kelvin-Helmholtz circulation is
well-conserved, in addition to rest-mass and total mass-energy,
linear and angular momentum.   Conservation of circulation indicates
that our code is not  seriously affected by numerical viscosity.  We
determine the  amplitude and frequency of the quasi-periodic
gravitational  waves emitted during the bar formation process using
the  quadrupole formula.
\end{abstract}

Stars in nature are usually rotating and subject to nonaxisymmetric
rotational instabilities.  An exact treatment of these instabilities
exists only for incompressible equilibrium fluids in Newtonian gravity.
For these configurations, global rotational instabilities arise from
non-radial toroidal modes when $\beta\equiv T/W$ exceeds a certain
critical value. Here $T$ and $W$ are the rotational kinetic and
gravitational binding energies.  There exist two different mechanisms
and corresponding timescales for bar-mode instabilities.  Uniformly
rotating, incompressible stars in Newtonian theory are {\em secularly}
unstable to bar-mode formation when $\beta \geq \beta_{\rm sec}
\simeq 0.14$.  This instability can grow only in the presence
of some dissipative mechanism, like viscosity or gravitational
radiation.  The growth time is determined by the dissipative
timescale, which is usually much longer than the dynamical timescale
of the system.  By contrast, a {\em dynamical} instability to bar-mode
formation occurs when $\beta \geq \beta_{\rm dyn} \simeq 0.27$. 
This instability is independent of any dissipative mechanisms, and the
growth time is the hydrodynamic timescale of the system.

Determining the onset of the dynamical bar-mode instability, as well
as the subsequent evolution of an unstable star, requires a fully
nonlinear hydrodynamic simulation.  Recently, simulations in
Newtonian theory \cite{TH,PDD} have shown that a higher degree of
differential rotation enhances the onset of dynamical instability. 
Simulations in full general relativity \cite{SBS} suggest that
nonlinear gravitation has a similar effect.

The purpose of post-Newtonian simulations \cite{SSBS} is twofold.  
We verify that  relativistic
gravitation alone {\it enhances} the dynamical instability, i.e.
$\beta_{\rm dyn}$ decreases with increasing compaction.  We also
show that in unstable configurations, the bar persists for at least
several rotational periods \cite{NCT,Brown} so that unstable stars are
promising sources of quasi-periodic gravitational waves.

\begin{figure}[ht] 
\centerline{\epsfig{file=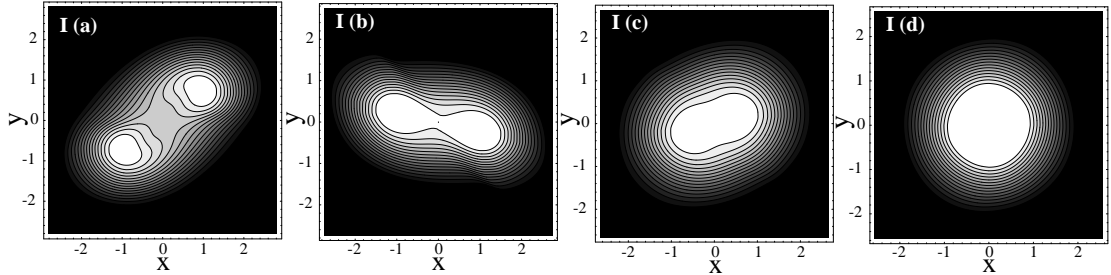,height=3.7cm,width=14.8cm}}
\vspace{10pt}
\caption{Final density contours in the
equatorial plane for differentially rotating stars all of compaction
$M/R = 0.05$ and varying values of $\beta$.  The contour lines
denote densities $\rho_{*}=1.3~i \times 10^{-3}$ ($i=1, \ldots, 15$).
Snapshots are plotted at the following times and initial $\beta$: 
(a) $t/P_{c}=2.72$,
$\beta = 0.265$ (b) $t/P_{c}=3.66$, $\beta = 0.259$, (c)
$t/P_{c}=7.77$, $\beta = 0.249$, and (d) $t/P_{c}=8.16$, $\beta =
0.238$.  Here, $P_{c}$ is the central rotation period.}
\label{figure1}
\end{figure}

\begin{figure}[ht] 
\centerline{\epsfig{file=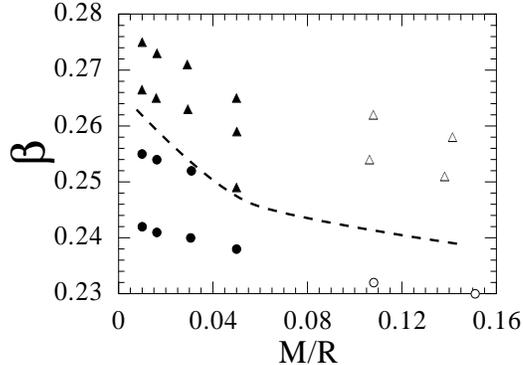,height=5cm,width=7cm}}
\vspace{10pt}
\caption{Summary of our dynamical stability analysis.  All our models are
plotted in a $\beta$ versus $M/R$ plane, with stable stars denoted by 
circles and unstable stars by triangles.  The solid circles and triangles
are the models studied in our post-Newtonian simulations 
{\protect \cite{SSBS}}; 
the open circles and triangles are
the models explored in full general relativity {\protect \cite{SBS}}.  
We conclude that the critical value of {\protect $\beta =
\beta_{\rm dyn}$} slightly decreases with increasing compaction
$M/R$.  This trend is emphasized by the dotted line, which 
approximates {\protect $\beta_{\rm dyn}$} as a function of $M/R$.}
\label{figure2}
\end{figure}

We perform post-Newtonian simulations of rapidly and differentially
rotating stars to investigate general relativistic effects on the
dynamical bar-mode instability for small compactions $M/R \leq
0.05$.   As a criterion for stability, we checked whether the distortion
parameter  follows an exponential growth.  The formation of a bar is also 
apparent in the
snapshots of density contours (See Fig. \ref{figure1} for $M/R = 0.05$; 
Figs. 4 -- 7 of  Ref. \cite{SSBS}). 
These plots clearly exhibit  a triaxial structure for the unstable
models, while for stable models  the density distribution hardly
changes during the evolution.  By combining these post-Newtonian
results with the  fully relativistic simulations \cite{SBS} for
configurations of higher compaction, we conclude that the critical
value of $\beta = \beta_{\rm dyn}$ decreases with increasing $M/R$
(Fig. \ref{figure2}).  Thus, relativistic gravitation enhances the
bar-mode instability.

We also confirm that bar-structure persists at least several rotation
period by checking the conservation of circulation (Fig. 8 and Table 3 
of Ref. \cite{SSBS}).  In the presence of significant numerical
viscosity, long-time evolution calculations become very unreliable and
may lead, for example, to erroneous evolution of a saturated bar.  We
have shown that in our calculations the circulation is well conserved,
implying that our code is at most very weakly affected by numerical
viscosity.  We present a method for computing the circulation which
can be applied in Newtonian, post-Newtonian and fully relativistic
calculations.

Finally, we have calculated  approximate gravitational waveforms in
the quadrupole approximation, neglecting all post-Newtonian corrections
(Fig. 10 of Ref. \cite{SSBS}).  We found that for unstable stars a
quasi-periodic oscillation with growing amplitude arises during the
early bar formation.  The  bar-structure persists for several rotation
periods, implying that bar-unstable stars are promising sources of 
quasi-periodic gravitational waves.



\begin{references}
\bibitem{TH}
Tohline, J. E. and Hachisu, I., 
{\it Astrophys. J.}\ {\bf 361}, 394 (1990).

\bibitem{PDD}
Pickett, B. K., Durisen, R. H., and Davis, G. A., 
{\it Astrophys. J.}\ {\bf 458}, 714 (1996).

\bibitem{SBS}
Shibata, M., Baumgarte, T. W., and Shapiro, S. L., {\it
Astrophys. J.}\ {\bf 542}, 453 (2000).

\bibitem{SSBS}
Saijo, M., Shibata, M., Baumgarte, T. W., and Shapiro, S. L., {\it
Astrophys. J.}\ {\bf 548}, 919 (2001).

\bibitem{NCT}
New, K. C. B., Centrella, J. M., and Tohline, J. E.,  
{\it Phys. Rev. D}\ {\bf 62}, 064019 (2000).

\bibitem{Brown}
Brown, J. D., 
{\it Phys. Rev. D}\ {\bf 62}, 084024 (2000).
\end{references}
\end{document}